\documentclass[aps,prl,twocolumn,superscriptaddress,showpacs]{revtex4}

\usepackage{color}
\usepackage{eso-pic}
\usepackage{amsmath}
\usepackage{tabularx}
\usepackage{booktabs}
\usepackage{textcomp}
\usepackage{rotating}
\usepackage{MnSymbol}
\usepackage{multirow}
\usepackage{graphicx}
\usepackage{lineno}
\newcommand{\abs}[1]{\lvert#1\rvert}
\newcommand{\met}       {\hbox{\textit{E}\kern-0.5em\lower-0.1ex\hbox{/}}_T}
\begin{document}

\title{Search for a low mass Standard Model Higgs boson in the $\tau\tau$ decay channel in $p\bar{p}$ collisions at $\sqrt{s}$ = 1.96 TeV}

\affiliation{Institute of Physics, Academia Sinica, Taipei, Taiwan 11529, Republic of China} 
\affiliation{Argonne National Laboratory, Argonne, Illinois 60439, USA} 
\affiliation{University of Athens, 157 71 Athens, Greece} 
\affiliation{Institut de Fisica d'Altes Energies, ICREA, Universitat Autonoma de Barcelona, E-08193, Bellaterra (Barcelona), Spain} 
\affiliation{Baylor University, Waco, Texas 76798, USA} 
\affiliation{Istituto Nazionale di Fisica Nucleare Bologna, $^{aa}$University of Bologna, I-40127 Bologna, Italy} 
\affiliation{University of California, Davis, Davis, California 95616, USA} 
\affiliation{University of California, Los Angeles, Los Angeles, California 90024, USA} 
\affiliation{Instituto de Fisica de Cantabria, CSIC-University of Cantabria, 39005 Santander, Spain} 
\affiliation{Carnegie Mellon University, Pittsburgh, Pennsylvania 15213, USA} 
\affiliation{Enrico Fermi Institute, University of Chicago, Chicago, Illinois 60637, USA}
\affiliation{Comenius University, 842 48 Bratislava, Slovakia; Institute of Experimental Physics, 040 01 Kosice, Slovakia} 
\affiliation{Joint Institute for Nuclear Research, RU-141980 Dubna, Russia} 
\affiliation{Duke University, Durham, North Carolina 27708, USA} 
\affiliation{Fermi National Accelerator Laboratory, Batavia, Illinois 60510, USA} 
\affiliation{University of Florida, Gainesville, Florida 32611, USA} 
\affiliation{Laboratori Nazionali di Frascati, Istituto Nazionale di Fisica Nucleare, I-00044 Frascati, Italy} 
\affiliation{University of Geneva, CH-1211 Geneva 4, Switzerland} 
\affiliation{Glasgow University, Glasgow G12 8QQ, United Kingdom} 
\affiliation{Harvard University, Cambridge, Massachusetts 02138, USA} 
\affiliation{Division of High Energy Physics, Department of Physics, University of Helsinki and Helsinki Institute of Physics, FIN-00014, Helsinki, Finland}\affiliation{University of Illinois, Urbana, Illinois 61801, USA} 
\affiliation{The Johns Hopkins University, Baltimore, Maryland 21218, USA} 
\affiliation{Institut f\"{u}r Experimentelle Kernphysik, Karlsruhe Institute of Technology, D-76131 Karlsruhe, Germany} 
\affiliation{Center for High Energy Physics: Kyungpook National University, Daegu 702-701, Korea; Seoul National University, Seoul 151-742, Korea; Sungkyunkwan University, Suwon 440-746, Korea; Korea Institute of Science and Technology Information, Daejeon 305-806, Korea; Chonnam National University, Gwangju 500-757, Korea; Chonbuk National University, Jeonju 561-756, Korea} 
\affiliation{Ernest Orlando Lawrence Berkeley National Laboratory, Berkeley, California 94720, USA} 
\affiliation{University of Liverpool, Liverpool L69 7ZE, United Kingdom} 
\affiliation{University College London, London WC1E 6BT, United Kingdom} 
\affiliation{Centro de Investigaciones Energeticas Medioambientales y Tecnologicas, E-28040 Madrid, Spain} 
\affiliation{Massachusetts Institute of Technology, Cambridge, Massachusetts 02139, USA} 
\affiliation{Institute of Particle Physics: McGill University, Montr\'{e}al, Qu\'{e}bec, Canada H3A~2T8; Simon Fraser University, Burnaby, British Columbia, Canada V5A~1S6; University of Toronto, Toronto, Ontario, Canada M5S~1A7; and TRIUMF, Vancouver, British Columbia, Canada V6T~2A3} 
\affiliation{University of Michigan, Ann Arbor, Michigan 48109, USA} 
\affiliation{Michigan State University, East Lansing, Michigan 48824, USA}
\affiliation{Institution for Theoretical and Experimental Physics, ITEP, Moscow 117259, Russia}
\affiliation{University of New Mexico, Albuquerque, New Mexico 87131, USA} 
\affiliation{Northwestern University, Evanston, Illinois 60208, USA} 
\affiliation{The Ohio State University, Columbus, Ohio 43210, USA} 
\affiliation{Okayama University, Okayama 700-8530, Japan} 
\affiliation{Osaka City University, Osaka 588, Japan} 
\affiliation{University of Oxford, Oxford OX1 3RH, United Kingdom} 
\affiliation{Istituto Nazionale di Fisica Nucleare, Sezione di Padova-Trento, $^{bb}$University of Padova, I-35131 Padova, Italy} 
\affiliation{LPNHE, Universite Pierre et Marie Curie/IN2P3-CNRS, UMR7585, Paris, F-75252 France} 
\affiliation{University of Pennsylvania, Philadelphia, Pennsylvania 19104, USA}
\affiliation{Istituto Nazionale di Fisica Nucleare Pisa, $^{cc}$University of Pisa, $^{dd}$University of Siena and $^{ee}$Scuola Normale Superiore, I-56127 Pisa, Italy} 
\affiliation{University of Pittsburgh, Pittsburgh, Pennsylvania 15260, USA} 
\affiliation{Purdue University, West Lafayette, Indiana 47907, USA} 
\affiliation{University of Rochester, Rochester, New York 14627, USA} 
\affiliation{The Rockefeller University, New York, New York 10065, USA} 
\affiliation{Istituto Nazionale di Fisica Nucleare, Sezione di Roma 1, $^{ff}$Sapienza Universit\`{a} di Roma, I-00185 Roma, Italy} 
\affiliation{Rutgers University, Piscataway, New Jersey 08855, USA} 
\affiliation{Texas A\&M University, College Station, Texas 77843, USA} 
\affiliation{Istituto Nazionale di Fisica Nucleare Trieste/Udine, I-34100 Trieste, $^{gg}$University of Udine, I-33100 Udine, Italy} 
\affiliation{University of Tsukuba, Tsukuba, Ibaraki 305, Japan} 
\affiliation{Tufts University, Medford, Massachusetts 02155, USA} 
\affiliation{University of Virginia, Charlottesville, Virginia 22906, USA}
\affiliation{Waseda University, Tokyo 169, Japan} 
\affiliation{Wayne State University, Detroit, Michigan 48201, USA} 
\affiliation{University of Wisconsin, Madison, Wisconsin 53706, USA} 
\affiliation{Yale University, New Haven, Connecticut 06520, USA} 
\affiliation{TRIUMF, Vancouver, British Columbia, Canada V6T~2A3}

\author{T.~Aaltonen}
\affiliation{Division of High Energy Physics, Department of Physics, University of Helsinki and Helsinki Institute of Physics, FIN-00014, Helsinki, Finland}
\author{B.~\'{A}lvarez~Gonz\'{a}lez$^w$}
\affiliation{Instituto de Fisica de Cantabria, CSIC-University of Cantabria, 39005 Santander, Spain}
\author{S.~Amerio}
\affiliation{Istituto Nazionale di Fisica Nucleare, Sezione di Padova-Trento, $^{bb}$University of Padova, I-35131 Padova, Italy} 
\author{D.~Amidei}
\affiliation{University of Michigan, Ann Arbor, Michigan 48109, USA}
\author{A.~Anastassov}
\affiliation{Northwestern University, Evanston, Illinois 60208, USA}
\author{A.~Annovi}
\affiliation{Laboratori Nazionali di Frascati, Istituto Nazionale di Fisica Nucleare, I-00044 Frascati, Italy}
\author{J.~Antos}
\affiliation{Comenius University, 842 48 Bratislava, Slovakia; Institute of Experimental Physics, 040 01 Kosice, Slovakia}
\author{G.~Apollinari}
\affiliation{Fermi National Accelerator Laboratory, Batavia, Illinois 60510, USA}
\author{J.A.~Appel}
\affiliation{Fermi National Accelerator Laboratory, Batavia, Illinois 60510, USA}
\author{A.~Apresyan}
\affiliation{Purdue University, West Lafayette, Indiana 47907, USA}
% ...
\collaboration{CDF Collaboration\footnote{With visitors from $^a$University of MA Amherst,
Amherst, MA 01003, USA,
$^b$Istituto Nazionale di Fisica Nucleare, Sezione di Cagliari, 09042 Monserrato (Cagliari), Italy,
$^c$University of CA Irvine, Irvine, CA  92697, USA,
$^d$University of CA Santa Barbara, Santa Barbara, CA 93106, USA,
$^e$University of CA Santa Cruz, Santa Cruz, CA  95064, USA,
$^f$CERN,CH-1211 Geneva, Switzerland,
$^g$Cornell University, Ithaca, NY  14853, USA, 
$^h$University of Cyprus, Nicosia CY-1678, Cyprus, 
$^i$University College Dublin, Dublin 4, Ireland,
$^j$University of Fukui, Fukui City, Fukui Prefecture, Japan 910-0017,
$^k$Universidad Iberoamericana, Mexico D.F., Mexico,
$^l$Iowa State University, Ames, IA  50011, USA,
$^m$University of Iowa, Iowa City, IA  52242, USA,
$^n$Kinki University, Higashi-Osaka City, Japan 577-8502,
$^o$Kansas State University, Manhattan, KS 66506, USA,
$^p$University of Manchester, Manchester M13 9PL, United Kingdom,
$^q$Queen Mary, University of London, London, E1 4NS, United Kingdom,
$^r$University of Melbourne, Victoria 3010, Australia,
$^s$Muons, Inc., Batavia, IL 60510, USA,
$^t$Nagasaki Institute of Applied Science, Nagasaki, Japan, 
$^u$National Research Nuclear University, Moscow, Russia,
$^v$University of Notre Dame, Notre Dame, IN 46556, USA,
$^w$Universidad de Oviedo, E-33007 Oviedo, Spain, 
$^x$Texas Tech University, Lubbock, TX  79609, USA,
$^y$Universidad Tecnica Federico Santa Maria, 110v Valparaiso, Chile,
$^z$Yarmouk University, Irbid 211-63, Jordan,
$^{hh}$University of Pennsylvania, Philadelphia, Pennsylvania 19104, USA,
$^{ii}$On leave from J.~Stefan Institute, Ljubljana, Slovenia, 
}}
\noaffiliation

\date{\today}

\begin{abstract}

We report on a search for the standard model Higgs boson decaying into pairs of $\tau$ leptons in                                                     
$p\bar{p}$ collisions produced by the Tevatron at $\sqrt{s}$ = 1.96 TeV. The analyzed data sample was                                                 
recorded by the CDFII detector and corresponds to an integrated luminosity of 6.0 fb$^{-1}$.         
The search is performed in the final state with one $\tau$ decaying leptonically and the second
one identified through its semi-hadronic decay.   
Since no significant excess is observed, a 95\% credibility level upper limit on the production cross 
section times branching ratio to the $\tau\tau$ final state is set for hypothetical Higgs boson masses between
 100 and 150 GeV/$c^2$. For a Higgs boson of 120 GeV/$c^2$ the observed (expected) limit is 14.6 (15.3)
 the predicted value.

\end{abstract}


\maketitle

Spontaneous electroweak symmetry breaking (EWSB) has been introduced in the standard model (SM) to explain how fundamental 
particles acquire mass. In the SM, EWSB requires the existence of a yet unobserved scalar particle, the Higgs boson~\cite{Higgs,Higgs_2,Higgs_3}. 
Direct searches at LEP~\cite{LEP}, combined with the recent exclusion mass ranges provided by the Tevatron~\cite{TEVATRON_2011_2} 
and LHC experiments~\cite{LHC}, have reduced the possible values for the mass of the SM Higgs boson. 
The Higgs mass is now excluded at 95\% confidence level (C.L.) below 114 GeV/$c^2$ and between 141 and 476 GeV/$c^2$. 
A global fit to electroweak measurements~\cite{globalfit} provides indirect indications that confirm the region below 158 GeV/$c^2$ 
as the most probable. Therefore the searches in the low mass decay channels play a crucial role to set a final statement on 
the SM Higgs boson existence.

Although the Higgs boson decay into a pair of $\tau$ leptons is not the dominant process in the low mass regime, searches in 
this final state are important for several reasons. First, the inclusion of additional channels improves overall sensitivity 
of the search. Also, once the Higgs boson is discovered, measurements of its decay branching ratios (B.R.'s) into different 
channels can give important information about its true nature. Furthermore, when beyond SM scenarios are considered, such the 
minimal supersymmetric extension of the SM (MSSM)~\cite{MSSM}, the study of final states containing $\tau$ leptons attracts even 
more interest, given their enhanced coupling to the Higgs bosons in large regions of the model parameter space.

A previous search for the SM Higgs boson in the final states with $\tau$ leptons was performed by the D0 Collaboration~\cite{D0-tau}. 
This Letter presents results of a search for the $H$ $\to$ $\tau\tau$ decays in the final state with two $\tau$ leptons and one or 
more jets, using a data sample corresponding to an integrated luminosity of 6.0 fb$^{-1}$ of $p\bar{p}$ collisions collected at the
 Tevatron by the CDFII detector at $\sqrt{s}$ = 1.96 TeV. The search is sensitive to contributions of several Higgs boson production 
mechanisms: gluon-gluon fusion $gg$ $\to$ $H$, associated production with the $W$ or $Z$ boson (VH), and the vector boson fusion (VBF)
 $qHq'$ $\to$ $q \tau\tau q'$.
To improve the search sensitivity a novel strategy for the $\tau$ identification is proposed, based on a multivariate technique~\cite{BDT3}.
Tau leptons are short-lived particles ($\tau$ = 290.6 $\times$ 10$^{\minus 15}$ s~\cite{bayesian}), which can be detected only through
their decay products. Semi-hadronic decays $\tau$ $\to$ $X_h \nu_{\tau}$, where $X_h$ is a system of hadrons and $\nu_{\tau}$ is the 
$\tau$ neutrino, are denoted in this paper as $\tau_h$ and correspond to a B.R. of about 65\%.                                        
With leptonic decays $\tau_l$ $\to$ $l \nu_l\nu_{\tau}$ accounting for the remaining 35\%, the final states $\tau_h\tau_l$, studied 
in this paper include about 46\% of all possible decays of $\tau$ pairs. 

 CDFII~\cite{cdf_a,cdf_b,cdf_c} is an azimuthally and forward-backward symmetric general-purpose particle detector. 
The geometry is described using the azimuthal angle  $\varphi$ and the pseudo-rapidity $\eta$ = $-\ln{\tan(\theta/2)}$, 
where $\theta$ is the polar angle measured with respect to the proton beam axis $z$. A charged-particle tracking system, 
immersed in a 1.4 T solenoidal magnetic field, consists of an inner set of silicon micro-strip detectors ($\abs{\eta} <$ 2),
and a open-cell drift chamber ($\abs{\eta} <$ 1). Electromagnetic and hadronic sampling calorimeters surround the tracking system
and cover central ($\abs{\eta}<$ 1.1) and forward (1.1 $< \abs{\eta} <$ 3.6) regions. Proportional chambers and scintillating 
strip detectors, embedded in the electromagnetic section of the central and forward calorimeters correspondingly, measure 
positions and transverse profiles of the showers. Dedicated muon detectors located outside the calorimeter provide coverage 
for $\abs{\eta} <$ 1.5. Gas Cherenkov counters, installed in the forward region 3.7 $< \abs{\eta} <$ 4.7, measure the average 
number of inelastic interactions per bunch crossing, providing a measurement of the instantaneous luminosity.
The components of energy and momenta in the plane transverse to the beam are defined as $E_T$ = $E\sin{\theta}$ 
and $p_T$ = $p\sin{\theta}$. Since neutrinos do not interact with the detector components, their $E_T$ must be inferred from the
energy imbalance in the transverse plane, $\met$~\cite{definitions}. 

Data for this search have been collected using a trigger selection~\cite{cdf_c} which requires one central electron or muon 
candidate with  p$_{ T} >$ 8 GeV/$c$ and one additional isolated track with p$_{ T} >$ 5 GeV/$c$~\cite{isolated_track}, which 
is used as a starting point for the $\tau_h$ reconstruction.

Jets are reconstructed from clustered energy depositions in the calorimeter towers using a fixed cone algorithm~\cite{JES} with a radius
in the $\varphi$-$\eta$ space $\Delta R$ = $\sqrt{\Delta\varphi^2+\Delta\eta^2} <$ 0.4. Jets are required to have $\abs{\eta}<$ 2.5, 
an electromagnetic fraction $E_{em}/E <$ 0.9, and $E_{T} >$ 20 GeV, where the transverse energy is corrected for instrumental effects 
and multiple $p\bar{p}$ collisions in the event~\cite{JES}.

At the analysis level, an electron candidate is reconstructed as a calorimeter cluster with E$_{ T} >$ 10 GeV pointed to by a
 well-measured track. The candidate is required to have a shower mostly contained in the electromagnetic compartment, and 
the lateral shape of the shower must be consistent with test beam measurements. A muon candidate is reconstructed from a track 
with $p_T >$ 10 GeV/$c$, matching to the track segments in the muon chambers. Muon energy deposition in the calorimeter should 
be consistent with that of a minimum-ionizing particle. Electron and muon reconstruction is described in detail in Ref.~\cite{cdf_c}.

In the detector, semi-hadronic decays of the $\tau$ leptons manifest themselves as narrow jets with low track multiplicity.
A $\tau_h$ candidate is required to have a good quality track with p$_{ T} >$ 6 GeV/$c$, pointing to a calorimeter cluster 
with $E_{ T}>$ 9 GeV and $\abs{\eta} <$ 1.0. Momenta of the charged and neutral particles reconstructed around the highest 
$p_T$ track in the cluster (referred to as the seed track~\cite{tau_cones}), define the \lq{}\lq{}visible\rq{}\rq{} 
four-momentum of the $\tau_h$ candidate, $\vec{p}_{vis}$. The transverse component of $\vec{p}_{vis}$, or visible $p_T$, 
is required to be greater than 15 GeV/$c$. Momenta of the neutral pions and photons are reconstructed using the central 
shower maximum detector. To be consistent with the topology of the hadronic $\tau$ decay, the number of tracks associated with 
the $\tau_h$ candidate is required to be 1 or 3. Tau decay modes with higher charged track multiplicity, contributing of the 
order of 1\%, are not considered. Contamination from electrons and muons mimicking the $\tau_h$ signature is reduced by rejecting 
candidates with $E_{had}/P < $ 0.2, where $E_{had}$ is the hadronic part of the $\tau_h$ cluster energy, and $P$ is the 
scalar sum of all track momenta associated with the $\tau_h$ candidate. 
To discriminate between taus and jets originating from quarks or gluons (referred to as QCD jets), the reconstructed $\tau_h$ 
candidates are required to be isolated. This is achieved by demanding that no charged or neutral particles are reconstructed 
in the isolation annulus~\cite{tau_cones} defined around the seed track, or by requiring the sum of their measured $p_T$ to be 
below a given threshold. In this search, the discrimination between $\tau_h$'s and QCD jets is further enhanced by using a 
multivariate selection based on a set of Boosted Decision Trees (BDTs)~\cite{BDT1,BDT2}.
Several $\tau_h$ categories are defined according to the number of tracks and to the visible $p_T$ of the $\tau_h$ candidate, 
and specific BDTs are implemented for each category. The input variables which provide the largest discrimination power are 
the $\tau_h$ visible mass, M$_{\mathrm vis}$ = $\sqrt{p_{\mathrm{vis}}^2}$, and the number of charged and neutral particles 
reconstructed in the isolation annulus, as well as the sum of their $p_T$. Compared to the cut-based $\tau_h$ selection~\cite{tauID}, 
the BDT-based selection reduces the jet $\to$ $\tau_h$ misidentification rate by 25\% while maintaining the same identification 
efficiency, at a level of 35\% for $p_T$ $>$ 30 GeV/$c$.

The two identified $\tau$ leptons in the event must have opposite charges and be spatially separated, $\Delta R >$ 0.4. In addition, 
selected events are required to have a good quality vertex with $\abs{V_z} <$ 60 cm, where $V_z$ is the vertex z-coordinate.

The irreducible background contributions to this search originate from $Z$+jets, $t\bar{t}$ and diboson events with one $\tau_h$ 
and an electron (or muon) in the final state. Another source of background events is due to lepton misidentification in $\gamma$+jet,
QCD multijet and $W$+jets production. $Z$ $\to$ $ee/\mu\mu$ events can satisfy the selection criteria when one of the two electrons 
or muons is mistakenly identified  as a $\tau_h$ candidate. This contribution is suppressed by vetoing events containing a $\tau_h$ with
$E_{had}/P <$ 0.4 and the dilepton invariant mass M$_{l\tau}$ consistent with that of the $Z$ boson.

 \begin{figure}                                                                                                                                           
 \includegraphics[width=80mm]{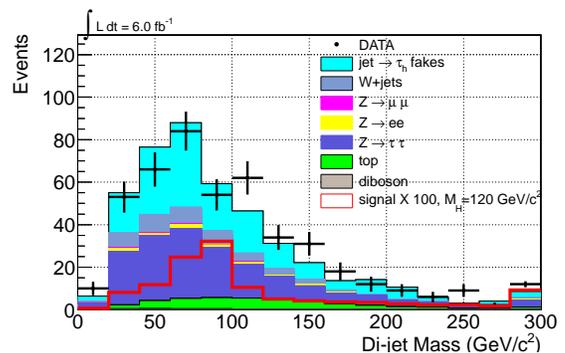}                                                                                                      
 \caption{\label{f:dijetmass}Invariant mass of the two most energetic jets, in the $N$(jets) $\ge$ 2 channel.}                                             
 \end{figure}                                                                                                                                          

The signal and background estimates are based on a combination of Monte Carlo (MC) simulations and data-driven methods. The detector 
response is simulated with a{ \sc geant}3-based package~\cite{GEANT}. Higgs, $t\bar{t}$ and diboson production processes are generated
by {\sc pythia}~\cite{PYTHIA} with the CTEQ5L set of parton distribution functions (PDF's)~\cite{PDF} and their contributions 
normalized to the next-to-leading order (NLO) theoretical cross sections~\cite{Campbell_Ellis_1, VH_uncertainty, ggH_uncertainty, Campbell_Ellis_2,Langenfeld}. The $Z$+jets background is generated using {\sc alpgen}~\cite{alpgen}, matched with {\sc pythia} for the hadronization and 
parton showering. Relative contributions of the different $Z$+$n$ partons subprocesses are provided by {\sc alpgen}, with the 
total event yield normalized  to the measured $Z$ $\to$ $ll$ production cross section~\cite{Ztautau}. Additional correction factors, 
determined by comparing data and simulations in specific control samples, are applied to account for the observed mis-modeling of the 
$e/\mu$ $\to$ $\tau_h$ misidentification probabilities. Background contributions resulting from lepton misidentification are estimated 
using a data-based method. In $\gamma$+jet and QCD multijet events no significant correlation is expected between the charges of the
two reconstructed lepton candidates, thus leading to an equal amount of opposite-sign (OS) and same-sign (SS) events with similar 
kinematic features. The $\gamma$+jet and QCD multijet contributions in the OS event selection are then modeled using the SS data sample.
In case of the $W$+jets process there is a correlation between the charge of the lepton produced in the $W$ boson decay and the 
charge of the outgoing quark which generates the jet misidentified as the $\tau_h$; this results in an excess of OS events.
This OS/SS asymmetry, A = 2.2 $\pm$ 0.5, is measured in a $W$+jets enriched control sample obtained by loosening 
the $\tau_h$ selection and by applying a cut on $\met$ $>$ 20/30/35 GeV (for events with $N$(jets) = 0, 1 or $\ge$ 2, respectively) and
 on the reconstructed $W$ transverse mass, M$_{ T}$ $>$ 40 GeV/$c$~\cite{definitions_2}.

We split the analysis sample into three subsamples according to the number of reconstructed jets in the final state. Events with 
$N$(jets) = 0 are used to validate the background model. The $\tau_h$ identification efficiency, $\epsilon$, is measured in a high 
purity $Z$ $\to$ $\tau\tau$ sample, defined by events with $N$(jets) = 0, $\met >$ 10 GeV and  M$_{T}$ $<$ 60 GeV/$c^2$. The ratio 
of the $\epsilon$ measurements in the data and MC, SF$_{\tau}$ = 0.96 $\pm$ 0.05, is used to correct the acceptance.
 Expected signal and background yields and the observed data in the $N$(jets) = 1 and $N$(jets) $\ge$ 2 channels are summarized 
in Table~\ref{t:signal}. Figure~\ref{f:dijetmass} shows the invariant mass of the two most energetic jets in the $N$(jets) $\ge$ 2 channel; 
since the hadronic decay of the $W$ or $Z$ boson is reconstructed, this variable is one of the most powerful in discriminating the associated 
Higgs boson production signal from the largest backgrounds.

\begin{table}
 \caption{\label{t:signal}Expected event yields in the two signal channels, compared to the number of observed events. 
Quoted errors include both systematic and statistical uncertainties.}
 \begin{ruledtabular}
 \begin{tabular}{lcc}
Process      & \multicolumn{1}{c} {$N$(jets) = 1} & \multicolumn{1}{c}{$N$(jets) $\ge$ 2} \\
\hline
$Z$ $\to$ $\tau \tau$          & 935 $\pm$ 85   & 160 $\pm$ 24\\
$Z$ $\to$ ee                  & 30.6 $\pm$ 3.9  & 6.8 $\pm$ 1.0\\
$Z$ $ \to$ $\mu \mu$          & 15.5 $\pm$ 2.7  & 1.9 $\pm$ 0.5\\
$WW$/$WZ$/$ZZ$                & 9.3 $\pm$ 1.0   & 2.2 $\pm$ 0.3\\
$t \overline t$               & 11.3 $\pm$ 1.6  & 39.3 $\pm$ 4.6\\
jet$\to \tau$ fakes           & 1219 $\pm$ 124  & 171 $\pm$ 19\\
$W$+jets                      & 151 $\pm$ 38    & 46.8 $\pm$ 14.1\\
\hline
Total background              & 2371 $\pm$ 156  & 428 $\pm$ 36\\
\hline
$gg$ $\rightarrow$ $H$        & 1.42 $\pm$ 0.45     & 0.34 $\pm$ 0.24 \\
$WH$                          & 0.23 $\pm$ 0.03     & 0.39 $\pm$ 0.05 \\
$ZH$                          & 0.13 $\pm$ 0.02     & 0.25 $\pm$ 0.03 \\
VBF                           & 0.18 $\pm$ 0.02     & 0.25 $\pm$ 0.03 \\
\hline
Total signal ($M_H$ = 120 GeV/c$^{2}$)& 1.96 $\pm$ 0.45     & 1.23 $\pm$ 0.29 \\     
\hline
Data                          & 2517              & 462 \\
 \end{tabular}
 \end{ruledtabular}
 \end{table} 

 Several sources of systematic uncertainties affect both the signal and background estimates. The largest one is the 
uncertainty on the jet energy scale~\cite{JES}, which leads to acceptance variations up to a 20\% level in different 
jet multiplicity channels. Non-negligible effects on the shapes of different jet-related kinematic distributions are 
also observed and taken into account. Another relevant source of systematic uncertainty which affects the sample 
normalizations is the theoretical and experimental uncertainty on the cross sections: uncertainties on diboson and 
$t \bar t$ production are 6\% and 10\% respectively~\cite{Campbell_Ellis_1, Campbell_Ellis_2,Langenfeld}, while for 
Drell-Yan processes we take a value of 2.2\%~\cite{Ztautau}. The $gg$ $\to$ $H$ production depends strongly on the 
gluon parton density function and the accompanying value of $\alpha_s$(q$^2$). The theoretical uncertainty on the 
cross section of this process, calculated separately for different jet multiplicity channels, results in uncertainties 
of 23.5\% and 67.5\% on the expected yield of the signal events with $N$(jets) = 1 and $N$(jets) $\ge$ 2, 
respectively~\cite{ggH_uncertainty}. The cross section uncertainties for the other Higgs boson production mechanisms 
are 10\% for VBF and 5\% for VH~\cite{Campbell_Ellis_1, VH_uncertainty}. The uncertainties on the $\tau_h$ identification 
efficiency and on the e $\to$ $\tau_h$ and $\mu$ $\to$ $\tau_h$ misidentification probabilities, are 5\%, 7\%, and 15\% 
respectively. Systematic uncertainties due to the modeling of the initial state (ISR) and final state radiation (FSR) have 
been evaluated for all the Higgs boson production processes: the largest variation, 15\%, corresponds to the gluon fusion 
process. The choice of the PDFs affects the acceptance at the level of a few percent, while an uncertainty of 2.3\% on 
the inclusive Drell-Yan acceptance is derived from the differences between the observed {\sc alpgen} and {\sc pythia}-based 
calculations. A 10\% uncertainty on the contribution of events with jets misidentified as $\tau_h$ is estimated from SS 
data. The uncertainty of the $W$+jets background is estimated, using OS/SS asymmetry, at 25\% (30\%) for $W$ + 1 ($\ge$ 2) 
jet channel.
\begin{figure*}
\includegraphics[width=80mm]{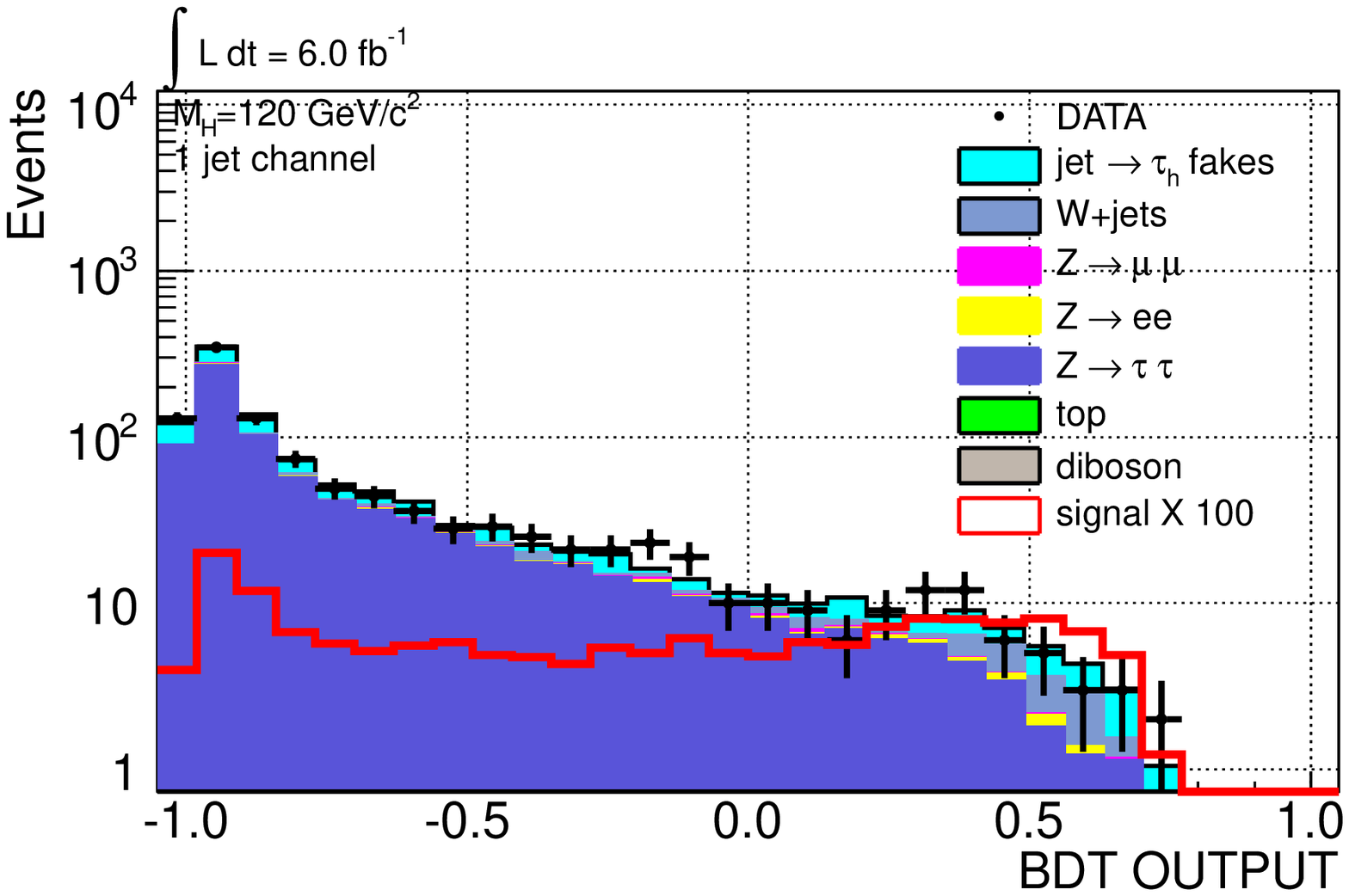}%
 \includegraphics[width=80mm]{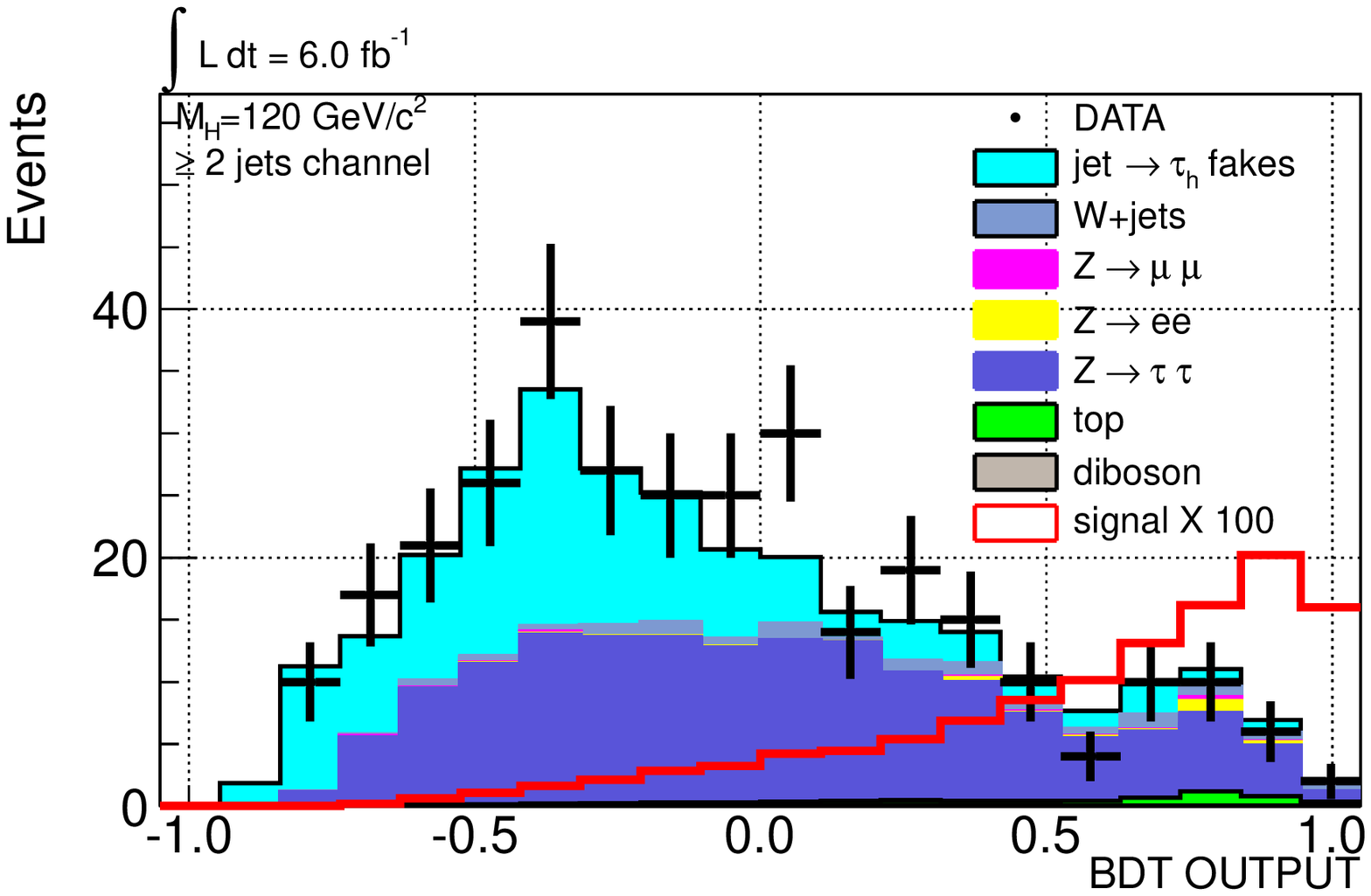}
 \caption{\label{BDT}Final discriminant distributions for the $N$(jets) = 1 and $N$(jets) $\ge$ 2 jets channels, obtained 
by combining a set of BDTs trained to discriminate a 120 GeV/$c^2$ Higgs boson signal from the main sources of background.}
 \end{figure*}
\begin{table*}
  \caption{\label{t:bayes_limit}Expected and observed 95\% C.L. upper limits on $\sigma(p\bar{p} \to H) B.R.(H \to \tau\tau)$, 
relative to the theoretical SM prediction.}
 \begin{ruledtabular}
 \begin{tabular}{lccccccccccc}
 $M_H$ (GeV/$c^2$) & 100 & 105 & 110 & 115 & 120 & 125 & 130 & 135 & 140 & 145 & 150 \\
 Expected &18.9&17.5&15.7&15.2&15.3&16.9&19.4&23.6&29.7&40.9&62.0\\
  Observed &19.6&18.1&15.5&14.7&14.6&16.4&18.7&22.6&33.7&47.3&70.2\\
 \end{tabular}
 \end{ruledtabular}
 \end{table*}

After all selection cuts are applied, the background expectation is still significantly larger than the expected Higgs
boson signal. The dominant irreducible background contribution, Z$\rightarrow \tau\tau$, is topologically very similar 
to the signal, thus affecting the sensitivity of the search; in addition, the Higgs boson invariant mass reconstructed 
from the visible decay products of the two $\tau$ leptons suffers from a very poor resolution, given that a sizeable 
fraction of the $\tau$ energy is carried by the undetected neutrinos. Since no other single powerful variable is available, 
a multivariate approach is followed and the discriminating power of various kinematic and topological observables is combined
into one single output discriminant. The search strategy is based on two sequential BDT-based selections applied to the 
candidate events. The first set of BDTs is trained to discriminate $H$ $\to$ $\tau\tau$ from $W$+jets, QCD, 
$Z$ $\to$ $ee/\mu\mu$ and $t\bar{t}$ backgrounds. At this step an event is accepted if its BDT score is higher than an 
optimal value, estimated by maximizing the search sensitivity. The second set of BDTs is trained to discriminate the signal 
against $Z$ $\to$ $\tau\tau$. The choice and the number of input variables is optimized for each BDT,  considering the most 
discriminating among a set of 23 well-modeled kinematic quantities. The training procedure is performed separately for the 
$N$(jets) = 1 and $N$(jets) $\ge$ 2 channels and optimized for two different Higgs boson mass scenarios 
($M_H$ = 120 GeV/$c^2$ and $M_H$ = 140 GeV/$c^2$). Results are shown in Fig.~\ref{BDT} for a Higgs boson mass of 120 GeV/$c^2$.

\begin{figure}[h!]
 \includegraphics[width=80mm]{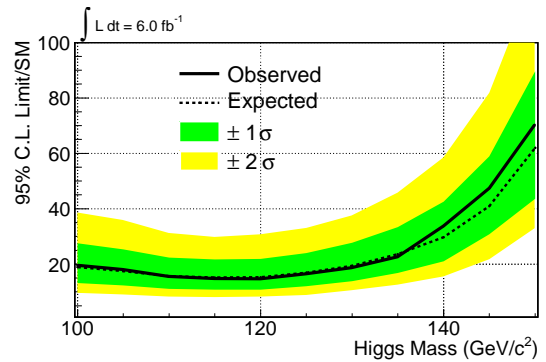}
 \caption{\label{f:bayes_limit}Expected (dashed curve) and observed (solid curve) 95\% C.L. upper limit on
$\sigma(p\bar{p} \to H) B(H \to \tau\tau)$, as a function of the mass, in the range between 100 and 150 GeV/$c^2$.}
 \end{figure}

The observed number of events in the two channels, as well as the distributions in the BDT output variables, are
consistent with the SM background expectations, so we use the BDT output templates to set 95\% C.L. upper limits 
on the SM Higgs boson production. A Bayesian likelihood method~\cite{bayesian} with a flat prior assigned to the 
signal cross section is employed, and all statistical and systematic shape and rate uncertainties are appropriately 
incorporated. The two channels are combined, and the resulting limits are shown in Fig.~\ref{f:bayes_limit} and 
summarized in Table~\ref{t:bayes_limit}. For all mass values, the expected and observed limits are within one 
standard deviation from each other.

We presented in this Letter the results of the search for the Higgs boson in the $\tau\tau$ decay mode, using final states involving
 one hadronically decaying $\tau$ lepton. This measurement contributes to the sensitivity of the low mass Higgs boson searches at the 
Tevatron. The value of the expected (observed) limit for a Higgs boson mass of 120 GeV/$c^2$ is 15.3 (14.6) times the SM expectation.

\begin{acknowledgments}
We thank the Fermilab staff and the technical staffs of the participating institutions for their vital contributions. This work was supported by the U.S. Department of Energy and National Science Foundation; the Italian Istituto Nazionale di Fisica Nucleare; the Ministry of Education, Culture, Sports, Science and Technology of Japan; the Natural Sciences and Engineering Research Council of Canada; the National Science Council of the Republic of China; the Swiss National Science Foundation; the A.P. Sloan Foundation; the Bundesministerium f\"ur Bildung und Forschung, Germany; the Korean World Class University Program, the National Research Foundation of Korea; the Science and Technology Facilities Council and the Royal Society, UK; the Institut National de Physique Nucleaire et Physique des Particules/CNRS; the Russian Foundation for Basic Research; the Ministerio de Ciencia e Innovaci\'{o}n, and Programa Consolider-Ingenio 2010, Spain; the Slovak R\&D Agency; the Academy of Finland; and the Australian Research Council (ARC). 
\end{acknowledgments}

\begin{thebibliography}{99}
\bibitem{definitions}The missing $E_T$ ($\met$) is defined as the magnitude of the vector $\vec{\met}$ = $-\sum_i E_T^i\hat{n_i}$, 
where $i$ is the calorimeter tower number with $\abs{\eta} <$ 3.6, while $\hat{n_i}$ is a unit vector perpendicular to the beam axis and pointing at the center of the i$^{th}$ calorimeter tower.
\bibitem{isolated_track}At trigger level, a track is isolated if no additional tracks with $p_T >$ 1.5 GeV/$c$ are reconstructed within 10 and 30 degrees around the track direction.
\bibitem{tau_cones}The $\tau_h$ reconstruction is based on a two-cone algorithm: the 3D angles $\theta_{sig}$ = min(0.17,5.0 GeV/$E_{clu}$) 
and $\theta_{iso}$ = 0.52 rad, where $E_{clu}$ is the cluster energy, define a signal cone and a isolation annulus around the highest $p_T$ track.
\bibitem{definitions_2}M$_{ T}$ = $\sqrt{2 p_T \met (1-\cos{\theta_{l\nu}})}$, where $p_T$ is the transverse momentum of the identified electron (or muon) in the final state and $\theta_{l\nu}$ is the azimuthal angle between the lepton and the direction of $\met$.
\end{thebibliography}

\end{document}